\documentclass[pss]{wiley2sp} 
\usepackage{amsmath}
\usepackage{color}

\definecolor{kg}{rgb}{1,0,0} 

\begin{document}

\title{Index assignment of a carbon nanotube rope using tip-enhanced Raman spectroscopy}

\titlerunning{Index assignment of a CNT rope using TERS}

\author{%
  Karin Go{\ss}\textsuperscript{\Ast,\textsf{\bfseries 1}},
  Niculina Peica\textsuperscript{\textsf{\bfseries 2}},
  Christian Thomsen\textsuperscript{\textsf{\bfseries 2}},
  Janina Maultzsch\textsuperscript{\textsf{\bfseries 2}},
  Claus M. Schneider\textsuperscript{\textsf{\bfseries 1}},
  Carola Meyer\textsuperscript{\textsf{\bfseries 1}}
  }

\authorrunning{Karin Go{\ss} et al.}

\mail{e-mail
  \textsf{k.goss@fz-juelich.de}, Phone:
  +49-2461-613485, Fax: +49-2461-612620}

\institute{%
  \textsuperscript{1}\,Peter Gr\"unberg Institut, Forschungszentrum J\"ulich \& JARA J\"ulich Aachen Research Alliance, 52425 J\"ulich, Germany\\
  \textsuperscript{2}\,Institut f\"ur Festk\"orperphysik, Technische Universit\"at Berlin, Hardenberg Str. 36, 10623 Berlin, Germany}

\received{XXXX, revised XXXX, accepted XXXX} 
\published{XXXX} 


\abstract{%
%
%
%
\abstcol{%
  We used tip-enhanced Raman spectroscopy to study the diameter-dependent Raman modes in a contacted carbon nanotube rope. We show that with the near-field tip enhancement a large number of nanotubes within a rope can be identified, even if the nanotube modes can not be
  }{%
  distinguished in the far-field signal. Several metallic and semiconducting nanotubes can be identified and assigned to nanotube families. Additionally, we provide a tentative chiral index assignment.}}

%
%

\maketitle   

\section{Introduction}
Raman spectroscopy probes the vibrational spectrum of a sample and thus reflects its molecular structure. For example, the spectrum of carbon nanotubes (CNTs) allows for determining the diameter, chiral index and electronic structure, if the resonance condition is fulfilled, i.\,e. the incident laser wavelength passes through an optical transition of the CNT \cite{Telg2004}. Recently, tip-enhanced Raman scattering (TERS) became the favourable spectroscopic method for detailed investigations of individual carbon nanotubes \cite{Hartschuh2003,Hartschuh2008}. Due to the excitation of localized surface plasmons at the apex of a metallic tip, the electromagnetic field is locally enhanced, resulting in a considerable increase of the Raman intensity. In addition, the diffraction limit for the spatial resolution is overcome, because the enhancement is localized to a few tens of nanometers in the vicinity of the tip. Considering these advantages, TERS is extremely promising as a characterization tool for nanoscale transport devices.

Here, we present TERS on a CNT rope with gold electrodes. The high spatial resolution of this method allows us to ensure that the Raman modes measured, i.e. the radial breathing mode (RBM) and the high energy mode (HEM), originate from CNTs of the transport device and not from adjacent CNTs that are not contacted. We identify several metallic and semiconducting CNTs, assign CNT families and provide a tentative chiral index assignment.

\begin{figure}[htb]
\includegraphics{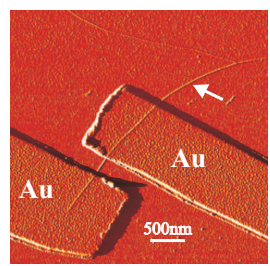}
\caption{Atomic force micrograph of the CNT rope device with Ti/Au contacts deposited on top of the CNTs. The tip position for the Raman measurements is indicated by an arrow.}
\label{fig:AFM}
\end{figure}

\section{Experimental details}

The CNTs of the device reported are grown on a $\mathrm{Si/SiO_2}$ substrate by chemical vapor deposition at 920$^\circ$C using an Fe/Mo catalyst and methane as precursor \cite{Kong1998}. At these temperatures, the process results mainly in single-walled CNTs and a few double-walled CNTs \cite{Spudat2009}. After the growth, contacts (5\,nm Ti/60\,nm Au) are patterned with a separation of 360\,nm by means of electron beam lithography. The height profile of an atomic force micrograph in Fig.~\ref{fig:AFM} shows a CNT rope height of $\sim$7\,nm. This value is found to be constant between the contacts and in the close vicinity of the contacted region.

Tip-enhanced Raman spectra are taken in a side-illumination configuration using an XE-100 (Park Systems) atomic force microscope (AFM) in contact mode combined with a LabRam HR-800 spectrometer (Horiba-Jobin-Yvon) with a spectral resolution of 2\,cm$^{-1}$. Available excitation energies of the laser setup are 2.33\,eV (double-frequency Nd-YAG, 532.2\,nm) and 1.96\,eV (He-Ne, 632.8\,nm). The $\mathrm{Si_3N_4}$ AFM tip coated with 80\,nm Au (tip radius $\sim$30\,nm) is positioned as marked in Fig.~\ref{fig:AFM} with a lateral resolution of $\sim$\,50-70\,nm. In the region between the two contacts it was not possible to approach the AFM tip towards the rope and obtain a CNT-related signal.

\section{Results}
\begin{figure}
\includegraphics{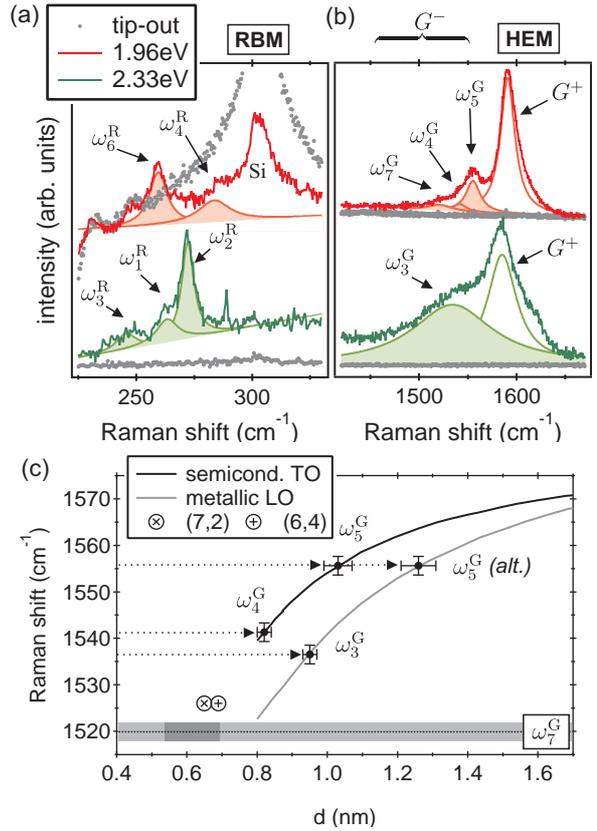}
\caption{Confocal (tip-out) and near-field (tip-in) Raman spectra at different excitation energies showing the region of (a)\,the RBMs and (b)\,the HEMs. (c)\,Diameter dependence of the $G^-$ mode for metallic and semiconducting CNTs \cite{Piscanec2007}. Observed $G^-$ frequencies are plotted and their diameter is extracted with an error defined by the instrumental resolution. For $\omega_5^{\mathrm{G}}$ an alternative diameter evaluation is drawn (see text for details). The experimentally observed $G^-$ frequency for a CNT assigned to either $(7,2)$ or $(6,4)$ is plotted.}
\label{fig:find-dia}
\end{figure}

Figures~\ref{fig:find-dia}a and b show representative confocal far-field (tip-out, grey) and near-field (tip-in, red and green) Raman spectra of the CNT rope. Besides the Si peak of the substrate ($\sim$\,302\,cm$^{-1}$), the confocal spectra show no Raman modes characteristic of a CNT in our experimental conditions (laser power and integration time).

In the near-field spectra, however, CNT-related Raman modes are clearly observable, i.\,e. RBMs, the defect-induced D mode (not shown), the HEM (split into two features: $G^-$ and $G^+$), and the 2D mode (not shown). For the identification of individual strands within the rope, we focus on the two diameter-dependent modes, i.\,e. the RBM and the $G^-$ mode. The frequencies of these modes remained constant for different positions of the AFM tip along the CNT rope, while their relative intensities changed.

In Fig.~\ref{fig:find-dia}a two RBMs $\omega_6^{\mathrm{R}}$ and $\omega_4^{\mathrm{R}}$ can be observed using an excitation energy of 1.96\,eV (see Tab.~\ref{tbl:assignment}). At an excitation energy of 2.33\,eV three different RBM frequencies ($\omega_3^{\mathrm{R}}$, $\omega_1^{\mathrm{R}}$ and $\omega_2^{\mathrm{R}}$) are found. According to the diameter dependence of the RBM frequency $\omega_i^{\mathrm{R}}=\frac{215\mathrm{cm}^{-1}\mathrm{nm}}{d_i}+18$\,$\mathrm{cm}^{-1}$ \cite{Telg2004}, the observed peaks correspond to CNTs of diameters $d_i=0.81 \ldots 0.95$\,nm, as given in Tab.~\ref{tbl:assignment}. Here, $i$ is the index of the specific CNT.

\begin{figure*}[tb]
\sidecaption
\includegraphics{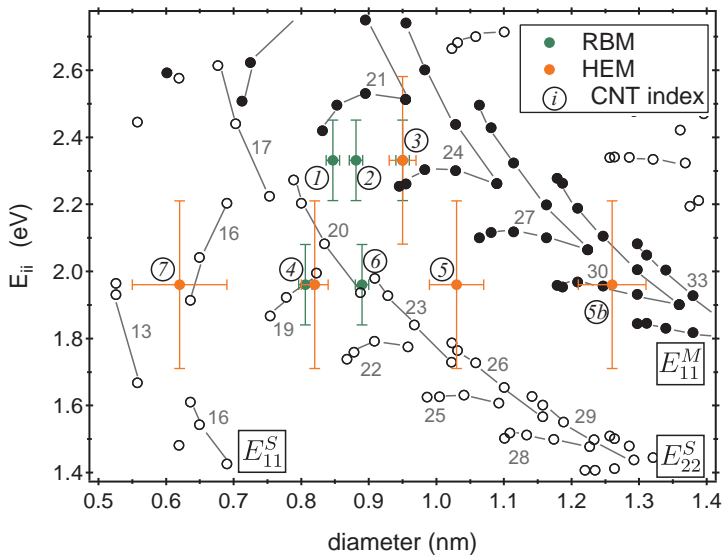}
\caption{Kataura plot as calculated by Popov \emph{et al.} \cite{Popov2004}. Full circles denote metallic CNTs, open circles denote semiconducting ones. Solid lines indicate nanotubes belonging to one CNT family, where the chiral index of a CNT is related to its left neighbour by $(n-1,m+2)$. Numbers give the specific family.}
\label{fig:Kataura}
\end{figure*}

Two characteristic HEM features are observed in Fig.~\ref{fig:find-dia}b: the $G^-$ and the $G^+$ peak. According to DFT calculations of Piscanec \emph{et al.} \cite{Piscanec2007} these peaks are ascribed to the transverse optical (TO) or the longitudinal optical (LO) phonon depending on the metallicity of the CNT. In the case of semiconducting CNTs, the $G^-$ peak is assigned to the TO phonon and the $G^+$ peak to the LO phonon, whereas in the case of metallic CNTs this assignment is vice versa.
In both cases the $G^-$ peaks exhibit a reasonable decrease of frequency with decreasing CNT diameter, as plotted in Fig.~\ref{fig:find-dia}c, whereas the $G^+$ peak is substantially diameter independent \cite{Piscanec2007,Jorio2002}. Depending on the chiral angle, sharp $G^-$ peaks (width $\sigma \sim$\,10\,cm$^{-1}$) \cite{Jorio2002} denote semiconducting CNTs, while for metallic CNTs the $G^-$ peak appears broadened ($\sigma \sim60$\,cm$^{-1}$) \cite{Oron-Carl2005} due to electron-phonon coupling \cite{Piscanec2007,Lazzeri2006}. Considering the linewidth of the $G^-$ peaks in Fig.~\ref{fig:find-dia}b we ascribe $\omega_3^{\mathrm{G}}$ to a metallic CNT ($\sigma = 102$\,cm$^{-1}$), and the three modes of the red spectrum ($\omega_7^{\mathrm{G}}$, $\omega_4^{\mathrm{G}}$ and $\omega_5^{\mathrm{G}}$) to semiconducting CNTs ($\sigma = 16\ldots30$\,cm$^{-1}$). The corresponding diameters of the CNTs are extracted from the diameter dependence of the $G^-$ peak in Fig.~\ref{fig:find-dia}c (see Tab.~\ref{tbl:assignment}). For CNT\,5, an alternative diameter for the case of a metallic CNT is given (see discussion below). Considering the spectrometer resolution we can deduce the error of the extracted CNT diameters as given by the error bars. However, $\omega_7^{\mathrm{G}}$ happens to be outside the calculated range. A previous experimental extension of the plot is added, where a semiconducting $G^-$ peak at 1526\,cm$^{-1}$ is assigned to the (7,2) or the (6,4) nanotube \cite{Telg2005}, thus allowing for an estimation of the respective CNT diameter ($\approx$\,0.62\,nm $\pm$\,0.07\,nm).

Only for CNT\,3 and CNT\,4 both the RBM and the $G^-$ mode can be observed. The RBM corresponding to $\omega_5^{\mathrm{G}}$ is for both assignments predicted at frequencies below the cut-off frequency of the spectrometer in the setup and thus not observable. The RBM of CNT\,7 ($\sim$376\,cm$^{-1}$) is expected in the region of the broad Si peak and is most likely hidden because of an expected low intensity (low intensity of $\omega_7^{\mathrm{G}}$).

The two missing $G^-$ peaks of CNT\,1 and CNT\,2 in the green spectrum are expected at similar Raman shifts ($\sim$1530\,cm$^{-1}$ and $\sim$1527\,cm$^{-1}$, respectively) close to $\omega_3^{\mathrm{G}}$ and with a broad linewidth due to the metallic character of these CNTs (discussed below). Considering these linewidths, the Lorentzian fits are unlikely to resolve the individual $G^-$ modes of the different diameters. Accordingly, the $G^-$ mode of CNT\,6 could easily be hidden by the several observed $G^-$ modes close to the expected Raman shift ($\sim$1548\,cm$^{-1}$).

Alltogether, we find modes corresponding to seven CNTs with different diameters that exclude the possibility of one large multi-walled CNT, considering a diameter difference of 0.68\,nm for nanotubes enclosing each other \cite{Kiang1998}.

\section{Discussion}

\begin{table*}[tb]
\sidecaption
\begin{tabular}{cccccc}
  \hline
  \hline
  $i$ & $\omega^{\mathrm{R}}$\,($\mathrm{cm^{-1}}$)   & $\omega^{\mathrm{G}}$($\mathrm{cm^{-1}}$) & $d$\,(nm) & $(2n+m)$ & ($n,m$) \\
  \hline
  1 & 262       &  --     & 0.88       & 21 & $(8,5)$\,$(9,3)$ \\
  2 & 272       &  --    & 0.85       & 21 & $(9,3)$\,$(10,1)$ \\
  3 & 245       &  1537  & 0.95     &   24    &  $(12,0)$\,$(11,2)$\,$(10,4)$ \\
  4 & 285       & 1541   & 0.81/0.82  & 19, 20 & $(8,4)$\,$(9,2)$\,$(7,5)$\\
  5 &    --    & 1556     & 1.26     &    30  &  $(13,4)$\,$(12,6)$\,$(11,8)$\\
  6 &260       &    --   & 0.89     & 20, 23 & $(7,6)$\,$(11,1)$ \\
  7 &             --        & 1520      & 0.62$\pm0.07$       &  16   & $(7,2)$\,$(6,4)$ \\
  \hline
  \hline
\end{tabular}
  \caption{All CNTs (index $i$) with the observed phonon frequencies and the corresponding diameters $d$ in nm. CNT families are assigned and possible chiral indices $(n,m)$ are given in the cases where it can be narrowed down to less than three alternatives.}
  \label{tbl:assignment}
\end{table*}

The observation of the CNT modes depends on the incident laser energy, which shows that the resonance effect plays a crucial role also in the tip-enhanced method of Raman spectroscopy. In the following we will ascribe the nanotubes, that we identified by our analysis, to so-called families or branches and make a tentative chiral index $(n,m)$ assignment considering a resonance effect.  An overview of the assignment is given in Tab.~\ref{tbl:assignment}. The Kataura plot in Fig.~\ref{fig:Kataura} shows optical transition energies $E_{ii}$ for $(n,m)$ CNTs versus their diameter, calculated within a non-orthogonal tight-binding model \cite{Popov2004}. Excitonic effects and electron correlations are taken into account by rigidly shifting the calculated values by 0.32\,eV \cite{Popov2004,Maultzsch2005}. Solid lines indicate nanotubes belonging to one CNT family with $(2n+m)=const$. The evaluated CNT diameters are drawn at the respective excitation energies. The typical width of the resonance window for RBMs in bundled CNTs is $\approx$\,120\,meV \cite{Fantini2004} depending on the diameter, while it is $\approx$\,250\,meV for the HEM due to the incoming and outgoing resonance being separated by the larger phonon energy of $\approx$\,200\,meV. Up to now, the influence of the tip enhancement on the resonance window is unknown \cite{Yano2006}. Hence, we assume large errors for $E_{ii}$ ($\pm$ the according resonance width), because a mode may be enhanced by several orders of magnitude in TERS, albeit being excited at the very edge of the resonance window. Indexing the observed CNT modes, an additional red-shift of the optical transitions for bundled CNTs has to be considered, whose magnitude was reported in the range of $50 \ldots 160$\,meV \cite{OConnell2004,Debarre2008,Michel2009,Fouquet2009}. This red-shift is attributed to a screening of Coulomb interactions inside the CNT due to its surrounding.


The modes observed for 2.33\,eV excitation energy can be clearly assigned to a specific CNT family. Considering a red-shift of the transition energies, CNT\,1 and CNT\,2 can unambiguously be assigned to the "21"-family ($(2n+m)=21$). Possible chiral indices for CNT\,1 are $(8,5)$ and $(9,3)$, while for CNT\,2 $(9,3)$ and $(10,1)$ are likely (note that $(9,3)$ can not be simultaneously assigned to both CNTs). In the case of CNT\,3 the "21"-family can be excluded, because for the only candidate, the $(7,7)$ armchair CNT, the LO phonon is forbidden by symmetry \cite{Reich2004}. Thus a $G^-$ mode would not be observable. Instead, we assign CNT\,3 to the adjacent "24"-family, where three chiral indices of the lower branch are possible.

The assignment of CNT\,4 to one specific family is not possible, although both of its modes are found in experiment. However, considering the red-shift mentioned above, we tentatively assign the nanotube to the "20"-family, rather than to the "19"-family. Out of the "20"-family $(9,2)$ or $(8,4)$ are candidates for CNT\,4, while $(7,5)$ out of the "19"-family can not be excluded.

The assignment for CNT\,5 is more difficult. A diameter of 1.03\,nm is found assuming $\omega_5^{\mathrm{G}}$ to be the TO phonon of a semiconducting CNT. In that case, the CNT has to be out of the lower $E_{22}^S$-branch of semiconducting CNTs, i.\,e. from the "23"- or the "26"-family. However, the intensity of the mode is higher than expected for such out-of-resonance CNTs, which might be caused by the high field enhancement at the AFM tip. An alternative route for assigning this mode is to assume it to be the LO phonon of a metallic CNT despite the small linewidth of 18\,$\mathrm{cm^{-1}}$. It has been shown, that the peak width for metallic CNTs can vary significantly as a function of the chiral structure \cite{Wu2007}. According to those results, $(n,0)$ CNTs exhibit the broadest $G^-$, and the broadening decreases for CNTs with increasing chiral angle until it is minimal for $(n,n)$ CNTs. Thus a narrow $G^-$ peak alone does not necessarily indicate a semiconducting CNT. Assuming a metallic CNT\,5, its diameter is extracted to be 1.26\,nm from Fig.~\ref{fig:find-dia}c. This alternative CNT diameter (labelled \emph{5b} in Fig.~\ref{fig:Kataura}) hits directly the $E_{11}^M$-branch of metallic tubes at the "30"-family. This latter assignment for CNT\,5 is favourable due to its clear consistency.

$\omega_6^{\mathrm{R}}$ originates either from a CNT within the "20"-family or from one within the neighbouring branch of the "23"-family ($(7,6)$ or $(11,1)$). From Raman spectroscopy alone it is not possible to gain more information about the chirality of this CNT. However, knowing the diameter precisely, as it could be determined by transmission electron microscopy (TEM), could succeed in an improved family assignment. A sample preparation technique offering the possibility to combine TEM with Raman measurements on contacted CNTs is presented by Frielinghaus et al. (in this volume).

Only the $G^-$ mode is observed for CNT\,7 and the fact, that the calculations in Fig.\,\ref{fig:find-dia}c do not include such low frequencies, leads to a large error for the diameter of this nanotube. Hence, despite the fact that the Kataura plot does not give many possiblities at such small diameters, the assignment is difficult. Considering once more the red-shift of transition energies due to bundling, an assignment to the "16"-family is reasonable. Within this family we can exclude the $(8,0)$ nanotube for symmetry reasons. The $(7,2)$ and the $(6,4)$ nanotube remain as possible assignments. However, a previously observed TO phonon frequency at a Raman shift 6\,$\mathrm{cm^{-1}}$ higher than the one here has been assigned to the same nanotubes (see Fig.~\ref{fig:find-dia}c) \cite{Telg2005}. This discrepancy in the $G^-$ frequency may be due to different environments, because the electron-phonon coupling is expected to be weakened for nanotubes enclosed by a surfactant and hence the Raman mode appears less down-shifted \cite{Telg2005}.

\section{Summary}
We found that, in our experimental conditions, the tip-enhancement of TERS is crucial to obtain Raman spectra of the contacted carbon nanotube rope. We identified seven individual nanotubes by their diameter-dependent phonons, the radial breathing mode and the high energy mode, excluding the possibility of a single multi-walled CNT. This is important, because the far-field signal might indicate a much lower number of nanotubes if only the modes exactly in resonance are observed. Assuming a broad resonance effect for the near-field spectra, we assigned each of the CNTs a nanotube family within the Kataura plot and found both metallic and semiconducting ones. Assigning the observed Raman modes to CNT families, we had to reconsider the metallicity of one of the CNTs, which changed the evaluated diameter reasonably. In this view, CNT\,5 could be the outer shell of a DWNT enclosing CNT\,7. The information on the metallicity and the chiral index of the involved CNTs gained by TERS can substantially support the interpretation of the electronic transport through the reported device \cite{Goss2011} and the local character of the spectroscopic method ensures a correct correlation.


\begin{acknowledgement}
The authors acknowledge S. Trellenkamp for e-beam writing and the DFG (FOR912) for funding. This work was supported by the Cluster of Excellence ''Unifying Concepts of Catalysis''.
\end{acknowledgement}

%
%
\providecommand{\WileyBibTextsc}{}
\let\textsc\WileyBibTextsc
\providecommand{\othercit}{}
\providecommand{\jr}[1]{#1}
\providecommand{\etal}{~et~al.}

\end{document}